\documentclass[prl,
twocolumn,
floatfix]{revtex4-1}
\usepackage{graphicx}
\usepackage{hyperref}
\usepackage[utf8]{inputenc}
\usepackage{color}

\usepackage{lmodern}
\usepackage{amsmath}
\usepackage{amssymb}
\usepackage{mathtools}
\usepackage{times}

\usepackage[english]{babel} 

\usepackage[tight,nice]{units}

\DeclareMathAlphabet{\bi}{OML}{cmm}{b}{it}

\newcommand{\diff}{\mathrm{d}}

\newcommand{\pabl}[2]{\frac{\partial #1}{\partial #2}}

\newcommand{\Ip}{I_{\mathrm{p}}}

\newcommand{\halb}{\frac{1}{2}}

\newcommand{\beq}{\begin{equation}}
\newcommand{\eeq}{\end{equation}}

\newcommand{\vxc}{v_\mathrm{xc}}

\newcommand{\vKS}{{v}_\mathrm{KS}}

\newcommand{\Wcmcm}{W/cm$^2$}

\begin{document}

\title{High-harmonic generation in solids with and without topological edge states}
\author{Dieter Bauer}
\affiliation{Institute of Physics, University of Rostock, 18051 Rostock, Germany}
\author{Kenneth K. Hansen}
\affiliation{Department of Physics and Astronomy, Aarhus University, DK-8000, Denmark}

\date{\today}

\begin{abstract} 
High-harmonic generation (HHG) in the two topological phases of a finite, one-dimensional, periodic structure is investigated using a self-consistent time-dependent density functional theory (TDDFT) approach. For harmonic photon energies smaller than the band gap, the harmonic yield is found to differ up to fourteen orders of magnitude for the two topological phases. This giant topological effect is explained by the degree of destructive interference in the harmonic emission of all valence-band (and edge-state) electrons, which strongly depends on whether topological edge states are present or not. The combination of strong-field laser physics with topological condensed matter opens up new possibilities to electronically control  strong-field-based light or particle sources or---vice versa---to steer by all optical means topological electronics.
\end{abstract}

\maketitle

{\em Introduction.} --- High-harmonic generation (HHG) in gases is one of the fundamental processes in intense laser-matter interaction. It paved the way for ``attosecond physics'' \cite{attoKrausz,attoCalegari}, and it is used to build compact short-wavelength sources for, e.g.,  single-shot imaging \cite{Rupp2017}. Condensed matter systems as targets require laser intensities below the damage threshold (unless one is interested in laser-plasma physics). However, recent experiments \cite{VampaPhysRevLett.115.193603,Ndabashimiye2016} and theoretical studies \cite{HawkinsIvanovPhysRevA.87.063842,Semiclassical_many_elec,gaarde_HHG,WuPhysRevA.94.063403,chinese_hhg_TDPI,ishikawa_HHG,Tancogne-Dejean2017,Hansen2017}  suggest that many of the strong-field concepts, like the three-step model of HHG \cite{RecollCorkumPhysRevLett.71.1994,LewensteinPhysRevA.49.2117}, seem to be applicable to condensed  matter as well, if appropriately adapted for the band structure and many-body effects \cite{Ghimire0953-4075-47-20-204030,VampaTutorial0953-4075-50-8-083001}. The existence of a band structure makes laser-solid interaction much richer (and complex) than laser-atom interaction in gases. In fact, while the essential target-dependent input for the strong-field approximation applied to HHG in atoms \cite{LewensteinPhysRevA.49.2117} is the ionization potential $\Ip$ and, as a preexponential correction, the transition matrix element between the initial state and a plane wave, the entire band structure matters in the case of solids. Due to the multiple conduction bands, multiple HHG plateaus are observed already at surprisingly low laser intensities \cite{Hansen2017}. The target-dependence implies that the band structure might be measurable by all-optical means \cite{VampaPhysRevLett.115.193603} and that the laser-solid interaction might be tunable for the benefit of useful applications based on light-driven electronics \cite{Schiffrin2013,Schultze1348,SchubertO.2014,Hassan2016,Sommer2016,Lucchini916,Garg2016,Higuchi2017}. 

As strong-field laser physics meets condensed matter, interdisciplinary aspects naturally come to the fore.  The laser may modify the band structure, creating ``Floquet matter'' \cite{Moessner2017,BucciPhysRevB.96.041126,Holthaus16,FaisalPhysRevA.56.748,DimiPhysRevA.95.063420}. In optical lattices, the laser even generates the band structure in the first place \cite{Bloch2005}. It is well known from condensed matter theory that---besides symmetry---the topology of the band structure plays a pivotal role in the understanding of, e.g., the spin Hall effect or topological insulators \cite{SpinHallRevModPhys.87.1213,topinsRevModPhys.82.3045,topins,topinsshortcourse}. Topological invariants allow to distinguish so-called topological phases, which are not only an end in themselves or useful for classification but also of practical interest because of their robustness with respect to imperfections in the samples, and potential applications such as topological superconductivity \cite{TopolinsandsupconRevModPhys.83.1057} or edge states in photonic Floquet topological insulators that are topologically protected from scattering \cite{Rechtsman2013}. 

We anticipate  that (relatively) intense,  short-pulse lasers will soon be pointed towards topological matter to record typical strong-field observables such as harmonics or photoelectron spectra. From the strong-field laser perspective, the question arises by which signatures topological effects may mani\-fest themselves in these observables. If there are direct links between, e.g., the HHG yield and the topology of the material one may either switch the strong-field observable electronically or control topological features by optical means, e.g., steering the spin currents along the edge states of a topological insulator on attosecond time scales. 

In this work, we investigate the influence of topological edge states on HHG spectra. To begin with, we focus on the simplest systems where such edge states appear, i.e., linear chains, which may serve as model systems for quasi-onedimensional systems  such as conjugated polymers, organic crystals, carbon nanotubes, ferromagnetic perovskites, carbon chains, transition metal complexes, or organic charge transfer salts  \cite{1dmetals,kagoshima2012one}. We will find a many-order-of-magnitude difference in the HHG yield between the two topologically different phases of linear chains and track its origin down to the different level of destructive interference with and without edge states.

\medskip

{\em Introduction of the model system.} --- Consider a linear chain of $N$ singly charged ions at positions $x_i$, separated by the lattice constant $a$,
\begin{align}
v_\mathrm{ion}(x) &=  - \sum_{i=1}^{N} \frac{1}{\sqrt{(x-x_i)^2 + 1}}, \\
 x_i  &=  \left[i- \frac{N+1}{2}\right]a. \label{eq:equidistant}
\end{align}
Here we employ the commonly used soft-core Coulomb potential for the interaction of electrons with ions in 1D  \cite{Eberly,BauerMacchiPhysRevA.68.033201,LeinKuemmelPhysRevLett.94.143003}, and atomic units $\hbar=|e|=m_e=4\pi\epsilon_0=1$ are used unless stated otherwise. We want to model the system self-consistently, beyond tight binding, with electron-electron interaction taken into account (at least on a mean-field level), and with the option to switch a laser on that is linearly polarized along the chain in order to study HHG. To that end we employ time-dependent density functional theory (TDDFT) \cite{RungeGross,UllrichBook} in  local spin-density approximation (LSD). The  time-dependent Kohn-Sham (KS) equation to be solved reads
\begin{align} 
i\partial_t  \varphi_{\sigma,i}(xt) & =  \left( - \frac{1}{2} \pabl{^2}{x^2} +  \vKS[\{n_{\sigma}\}](xt) \right)\varphi_{\sigma,i}(xt), \label{eq:KS}
\end{align}
with the KS potential 
\begin{align} \vKS[\{n_{\sigma}\}](xt) &= v_\mathrm{ion}(x)  - iA(t)\partial_x  \nonumber \\
  & \quad +  u[n](xt) + \vxc[\{n_{\sigma}\}](xt) \label{eq:KSpot} \end{align} where
$ u[n](xt) = \int n(x't)[(x-x')^2 + 1]^{-1/2}\, \diff x'$ is the Hartree potential,
$\vxc[\{n_{\sigma}\}](xt)  \simeq  -\left( \frac{6}{\pi} n_\sigma(xt)\right)^{1/3}$ is the exchange-correlation (xc) potential in x-only local spin-density approximation,  
$n_\sigma(xt) = \sum_{i=1}^{N_{\sigma}} |\varphi_{\sigma,i}(xt)|^2$ are the spin densities for spin $\sigma=\uparrow,\downarrow$, and $n(xt)=\sum_{\sigma} n_\sigma(xt)$ is the total single-particle density. The index $i$ in \eqref{eq:KS} runs over $i=1,2, \ldots, N_\sigma$ where $N_\sigma$ is the number of KS electrons of spin $\sigma$. We use the LSD exchange expression for the 3D electron gas because we mimic 3D electrons that are driven in the polarization direction of a linearly polarized laser rather than a true 1D electron system. The LSD correlation part is neglected, as it does not affect the qualitative findings that follow. We propagate the KS orbitals according  \eqref{eq:KS}, using a split-operator Crank-Nicolson approach for the time-evolution operator, and a predictor-corrector step to update the KS potential \cite{QSFQDBook}.

With the laser off, the vector potential vanishes,  $A(t) \equiv 0$, and the problem becomes stationary,
\begin{align} 
\epsilon_{\sigma,i}  \varphi_{\sigma,i}(x) & =  \left( - \frac{1}{2} \pabl{^2}{x^2} +  \vKS[\{n_{\sigma}\}](x) \right)\varphi_{\sigma,i}(x), \label{eq:tiKS}
\end{align}
with the KS orbital energies $\epsilon_{\sigma,i}$. The total energy of the model system is
$
E_\mathrm{tot} = E[\{ \varphi_{\sigma,i}\}] + E_\mathrm{ii} 
$
with the static ion-ion energy $E_\mathrm{ii}  = \sum_{i=1}^N\sum_{j<i} [(x_j-x_i)^2 + 1]^{-1/2}$
 and the electronic energy from the (TD)DFT simulation $E[\{ \varphi_{\sigma,i}\}] = T_s[\{ \varphi_{\sigma,i}\}] + E_\mathrm{ei}[n] + U[n] + E_\mathrm{xc}[n_\sigma]$ where
$ T_s[\{ \varphi_{\sigma,i}\} ] = -\halb\sum_\sigma\sum_{i=1}^{N_\sigma} \int\diff x\,  \varphi_{\sigma,i}^*(x) \pabl{^2}{x^2} \varphi_{\sigma,i}(x)$ is the kinetic energy, $ E_\mathrm{ei}[n] = \int v_\mathrm{ion}(x) n(x)\, \diff x$, $U[n] = \halb \int u[n](x) n(x)\, \diff x$, and
$ E_\mathrm{xc}[n_\sigma] \simeq \left( E_\mathrm{x}[2n_\uparrow] + E_\mathrm{x}[2n_\downarrow] \right)/2$ with $E_\mathrm{x}[n] =-\frac{3}{4} \left(\frac{3}{\pi}\right)^{1/3} \int n^{4/3}(x)\, \diff x$ \cite{dftparr}. The stationary KS ground-state orbitals are found via imaginary-time propagation in combination with Gram-Schmidt orthogonalization  \cite{QSFQDBook}. 

\medskip

{\em Appearance of topological edge states.} --- The model system undergoes a Peierls transition if the ion positions are modified according to
\begin{align}
x_i & \longrightarrow x'_i = x_i - (-1)^{i} \delta, \quad i=1,2,3, \ldots N, \label{eq:shift}
\end{align} 
i.e., alternatingly shifted to the left and to the right by $\delta$. 

\begin{figure}[h]
  \centering
  \includegraphics[width=0.9\columnwidth]{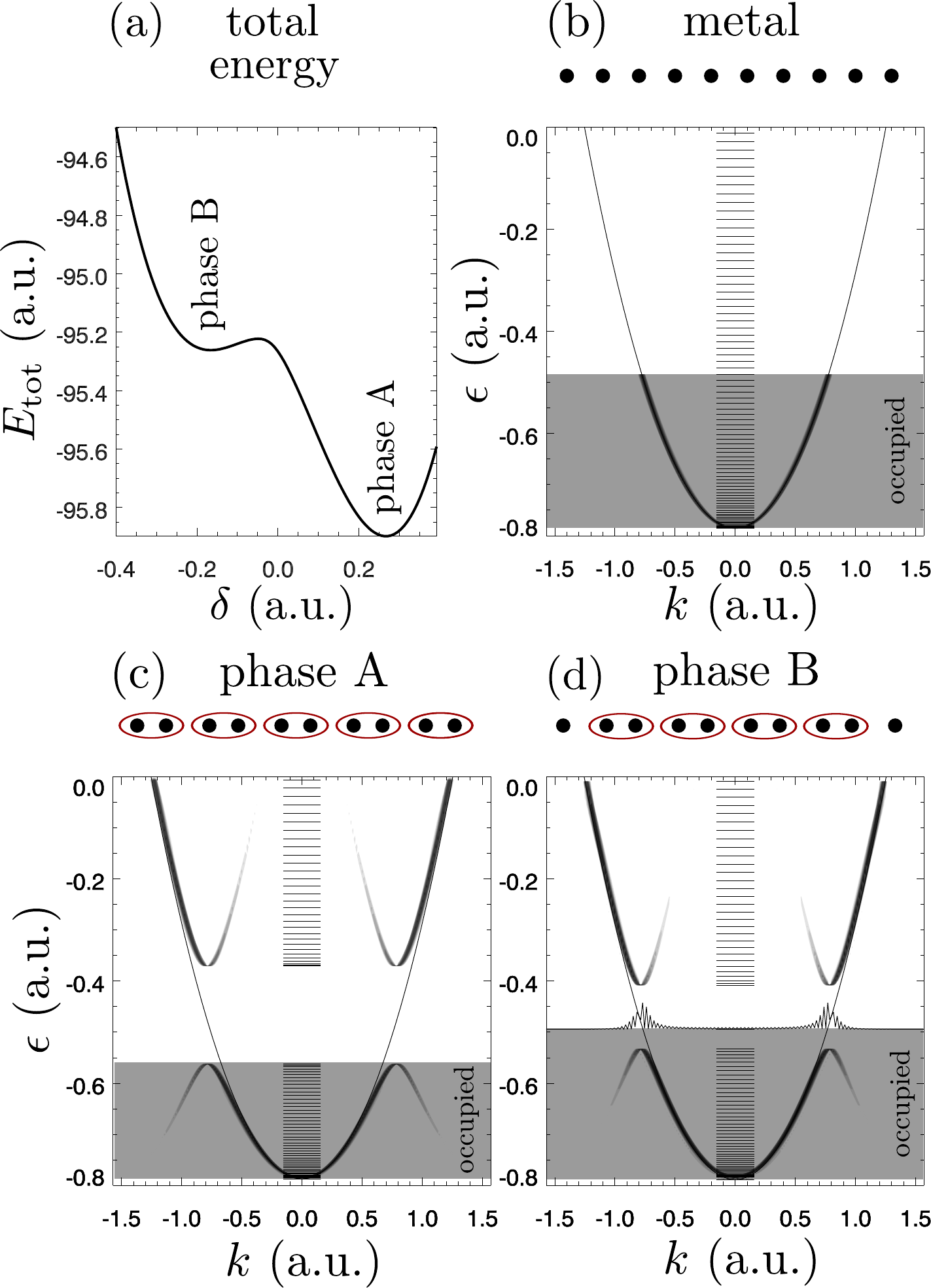}
  \caption{Features of the model system for $a=2$, $N=100$, and spin-neutral configuration $N_\uparrow=N_\downarrow=N/2$. (a) Total energy {\em vs} shift $\delta$ in Eq.~\eqref{eq:shift}. (b) Band structure for $\delta=0$ (metal), (c) $\delta=0.265$ (phase A), and (d) $\delta=-0.165$ (phase B). Occupied and unoccupied KS orbitals were Fourier-transformed from position to $k$-space, and the modulus square was plotted as a logarithmically scaled contour plot {\em vs} $k$ and orbital energy $\epsilon$. KS orbital energies $\epsilon_{\sigma,i}$ are indicated by horizontal bars. Gray backgrounds indicate the region of occupied levels. The shifts and the dimerization of the ions are illustrated by black dots and red ellipses, respectively, in panels (b), (c), (d). Further, the parabola $k^2/2+\epsilon_{\sigma,1}$ is drawn as a thin solid line in panels (b), (c), (d). The probability density of the occupied edge state in $k$-space is overplotted in (d) at the level of the corresponding KS orbital energy.  }
  \label{fig:bandstruc}
\end{figure}

Figure \ref{fig:bandstruc}(a) shows the total energy of the system as a function of the shift $\delta$ for lattice constant $a=2$, $N=100$ ions and electrons, and the spin-neutral configuration $N_\uparrow=N_\downarrow=N/2$. One observes an absolute minimum in energy at $\delta_A=0.265$ and a local minimum at $\delta_B=-0.165$. Figure~\ref{fig:bandstruc}(b)--(d) show the band structures for $\delta=0$, $\delta_A$, and $\delta_B$, respectively. Equidistant ions ($\delta=0$) lead to a half-filled band, i.e., a metal. However, panel (a) shows that this metallic phase $\delta=0$ is unstable (Peierls instability), resulting in a metal-to-insulator transition. This is because a finite $\delta$ implies that the lattice constant doubles to $2a$ because of the ``dimerization'' indicated by red ellipses in panels (c) and (d). This results in a bisection of the Brillouin zone from $[-\pi/a,\pi/a]$ in the metallic case with a half-filled lowest band to $[-\pi/2a,\pi/2a]$ for phase A and phase B with a fully occupied lowest band. Phase B has a smaller band gap than phase A, and, most importantly,  there are two extra states in the band gap due to the unpaired ions at the left and at the right edge of the chain. The lower of these almost degenerate edge states is occupied in the ground state KS configuration. Its $k$-space probability density  is indicated in panel (d), showing that it is less localized in $k$-space than the other, ``regular'' states. This is because edge states are rather localized at the edges in position space. 

The qualitative behavior of our model system concerning the Peierls transition and the appearance of edge states in phase B is similar to the Su-Schrieffer-Heeger  (SSH) model \cite{SSHPhysRevLett.42.1698,topins,topinsshortcourse}. However, we neither adopt a tight-binding approximation nor do we  restrict the model to two bands, break the  gauge invariance with respect to the coupling to the laser \cite{PhysRevB.66.165212}) or assume non-interacting or spin-less electrons. 

\medskip

{\em High-harmonics spectra.} --- We now let the chain interact with an $n_\mathrm{cyc}=5$-cycle  $\sin^2$-shaped laser pulse of frequency $\omega=0.0075$ (i.e., $\lambda\simeq 6.1\,\mu$m), described by the vector potential
\begin{align}
A(t) &= A_0 \sin^2\left(\frac{\omega t}{2 n_\mathrm{cyc}}\right) \sin\omega t
\end{align}
for  $0<t<n_\mathrm{cyc} {2\pi}/{\omega}$ (and zero otherwise).
The topological effect we discuss in the following is rather insensitive to the laser intensity as long as it is below the damage threshold but strong enough to generate high-order harmonics at all. In the simulations whose results are presented in the following, the vector potential amplitude was chosen $A_0=0.1$, corresponding to $\simeq 2 \times 10^{10}\,$\Wcmcm.

Besides the velocity-gauge coup\-ling $-iA(t)\partial_x$, the Hartree and the xc term in the KS potential \eqref{eq:KSpot} also become time-dependent, because they depend on the time-dependent (spin) density. The question is whether this time-dependence is important or not. As we must not destroy the solid with a too strong laser, the electron density has to stay close to the ground state density, and thus the KS potential close to the ground state KS potential. ``Freezing'' the KS potential to its ground state form corresponds to the simulation of $N$ non-interacting electrons in a given, static potential (for whose calculation interaction between the electrons was taken into account though). Instead, updating the KS potential each time step is a full TDDFT simulation with electron-electron interaction included on an LSD mean-field level. 
 
\begin{figure}[h]
  \centering
  \includegraphics[width=0.9\columnwidth]{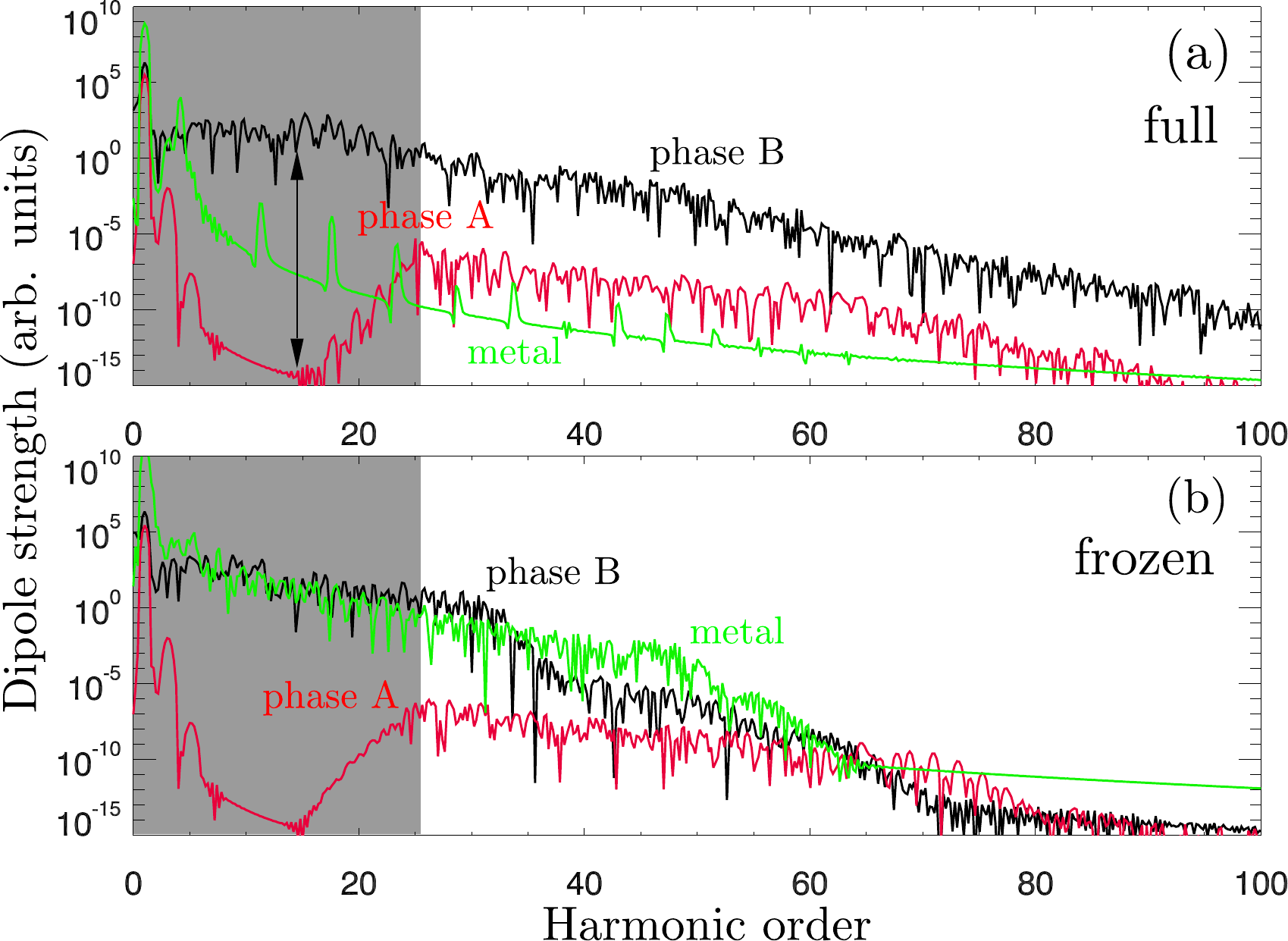}
  \caption{HHG spectra for phase A (without edge states) and phase B (with edge states) from the full TDDFT calculation (a) and a calculation with frozen ground-state KS potential (b). The results for the unstable metal phase are included for the sake of completeness. The gray-shaded areas indicate the sub-band-gap harmonics regime. The black double arrow in panel (a) highlights the many-order-of-magnitude enhancement in HHG efficiency in phase B compared to phase A.}
  \label{fig:spectra1}
\end{figure}

Figure~\ref{fig:spectra1} shows HHG spectra for the two phases A and B, and the metal. By the total ``dipole strength'' we understand $D(\omega) \propto |\mathrm{FFT}[X(t)]|^2$, i.e., the absolute square of the (Hanning-windowed) fast-Fourier-transform of  
\beq  X(t) = \sum_{i,\sigma} x_{\sigma,i}(t), \quad x_{\sigma,i}(t) = \int\diff x\, x|\varphi_{\sigma,i}(xt)|^2 . \eeq 
Alternatively, one may calculate the HHG spectra from the absolute square of the Fourier-transformed current or acceleration, giving essentially the same result apart from factors $\omega^2$ and $\omega^4$, respectively \cite{PhysRevA.79.023403,0953-4075-44-11-115601}. 

The results from the full TDDFT simulations are shown in Fig.~\ref{fig:spectra1}(a), from the frozen KS potential in Fig.~\ref{fig:spectra1}(b).  For phases A and B, the differences between the results for updated and frozen KS potential  are minor up to harmonic order $\simeq 30$. 
The gray-shaded area up to harmonic order 25 indicates the band gap between the valence and the first conduction band for phase A [see Fig.~\ref{fig:bandstruc}(c)]. The three-step model for solids \cite{VampaTutorial0953-4075-50-8-083001} predicts harmonics above the band gap because of the recombination step involving an electron in the conduction band and a hole in the valence band. The band gap thus plays the role of the ionization potential in conventional HHG in gases. Sub-band-gap harmonics can only be generated via the intraband motion of electrons in their (not perfectly parabolic) valence band. One could expect that in a fully occupied valence band the Pauli principle should prohibit such motion. Not so in the non-interacting KS system where each KS orbital moves independently and indeed generates strong sub-band-gap harmonics. Only due to destructive interference of all the dipoles of the individual KS electrons in a completely filled band almost no harmonics are emitted by phase A in the sub-band-gap area. The destructive interference occurs because roughly one half of the KS electrons with positive band curvature move oppositely to the other half with negative band curvature.   We see in Fig.~\ref{fig:spectra1} that only the 3rd and 5th harmonic survive for phase A. In any case, we checked explicitly that all KS orbitals stay orthogonal during time propagation, ensuring that the Pauli principle is always fulfilled.

The main result of this paper is the strong emission of harmonics by phase B below the band gap. In fact, the black double-arrow in Fig.~\ref{fig:spectra1}(a) indicates the 14-orders-of-magnitude topological effect we observe. The effect is present also for the frozen KS potential in Fig.~\ref{fig:spectra1}(b) and thus not due to electron-electron interaction. In that sense the many-orders-of-magnitude enhancement of the HHG efficiency is as robust as topological effects typically are,  and it will also not depend on the details of the xc-potential chosen in a TDDFT simulation. 

Although not relevant for the topic of this paper, for the sake of completeness, the HHG spectra for the metallic phase $\delta=0$ are included in Fig.~\ref{fig:spectra1}. A large difference between the spectra obtained with frozen and updated KS potential is observed for the metallic phase because screening due to the polarization of the metallic slab in the laser field is not taken into account when the KS potential is frozen \cite{kreibig}.

\begin{figure}[h]

  \centering
  \includegraphics[width=0.9\columnwidth]{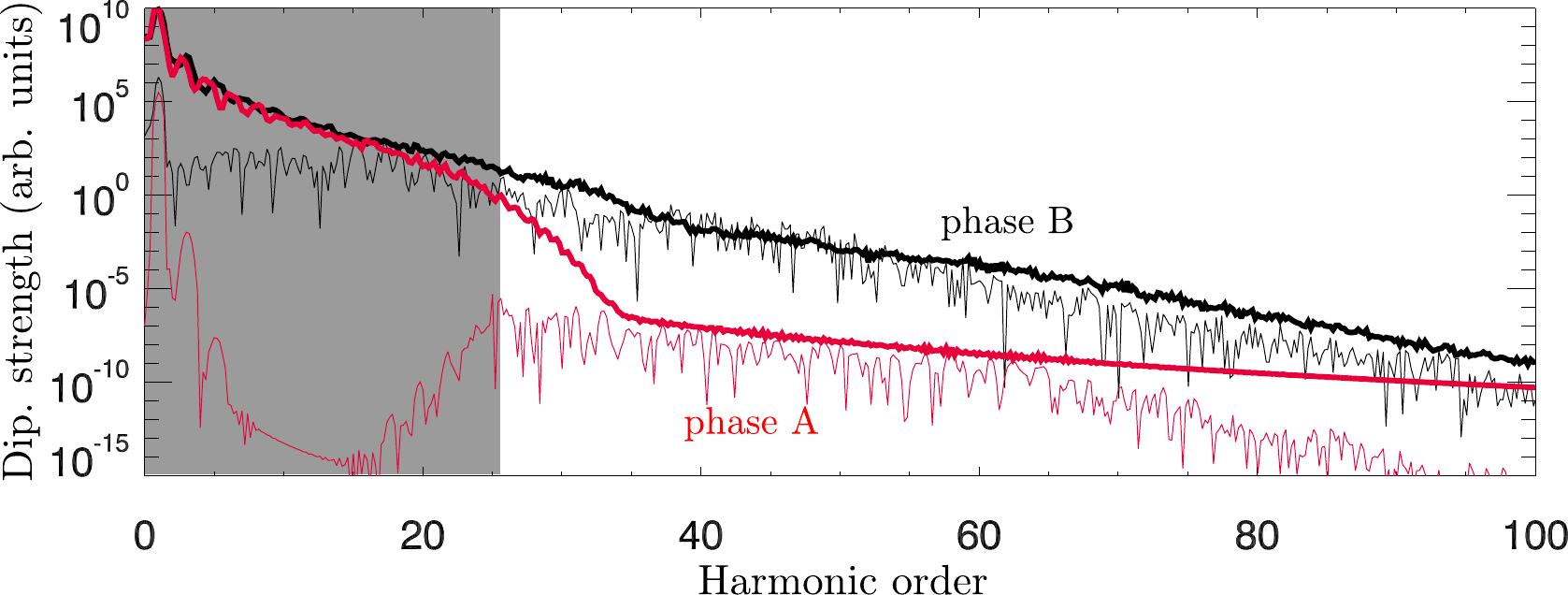}
  \caption{Incoherent sum of the individual, hypothetical spectra $d_{\sigma,i}(\omega) \propto |\mathrm{FFT}[x_{\sigma,i}(t)]|^2$ from the full-TDDFT simulations (bold). The true HHG spectra from Fig.~\ref{fig:spectra1}(a) are included (thin). While the incoherent sums are very similar for both phases in the sub-band-gap region (gray-shaded), the HHG spectra are many orders of magnitude different, highlighting the importance of almost complete and incomplete destructive interference for phase A and phase B, respectively.     }
  \label{fig:spectra2}
\end{figure}

The key question is why phase B produces high harmonics so much more efficiently in the sub-band-gap range than phase A. Let us consider hypothetical HHG spectra calculated as an {\em incoherent} sum over the individual KS spectra $d_{\sigma,i}(\omega) \propto |\mathrm{FFT}[x_{\sigma,i}(t)]|^2$. All individual spectra show strong harmonic emission within the band-gap, and so does their incoherent sum $\sum_{i,\sigma} d_{\sigma,i}(\omega)$, shown in Fig.~\ref{fig:spectra2}. The incoherent sum lies for almost all harmonic frequencies above the true HHG spectra, which proves that destructive interference is essential. While the incoherent sum is quite similar for both phases in the gray-shaded sub-band-gap region, the degree of destructive interference is many orders of magnitude different. In both phases the valence band is completely filled. However, in phase B edge states exist, and only  one of the two degenerate edge states is populated.  The two edge states in phase B thus act like a half-filled ``mini band,'' resulting in incomplete destructive interference during the emission of sub-band-gap harmonics.

{\em Topological vs other edge states.} --- Impurities may also modify the level structure or generate edge states in the band gap. Further, for an odd number of ions and one electron per ion, the chain is necessarily spin-polarized, and phase A and B become equivalent, just with an undimerized ion at the left or right boundary, respectively. In the Supplemental Material \cite{supplmat} the band structure and the HHG spectra for such cases are shown and discussed. In brief, (i) the edge state due to an odd number of ions does not lie in the band gap but below the valence band, and no HHG in the sub-band-gap region is observed. (ii) Two impurity ions at the left and right end of the chain in phase A generate two degenerate edge states in the band gap, which, however, are both occupied. As a result, destructive interference is more pronounced than in the pure phase-B case where only one of two degenerate edge states is occupied. (iii) Impurities at the left and right end of the phase-B chain merely shifts the two anyway existing degenerate edge states; efficient sub-band-gap HHG prevails (demonstrating the robustness of topological effects). (iv) Adding an additional, virtual KS electron that occupies the vacant edge state of phase B reduces the harmonic yield, confirming our claim that it is the incomplete destructive interference due to the half occupied edge states in phase B that leads to the many-order-of-magnitude stronger harmonic emission by phase B.

\bigskip

{\em Conclusion and outlook.} --- We found that the two topological phases of one-dimensional chains result in different strong-field harmonics spectra. Generation of sub-band-gap harmonics is orders of magnitude more efficient when half-filled topological edge states in the band gap are present. As we studied the simplest system that features topological edge states at all, our work is only a first step towards ``topological strong-field physics.'' In 1D, a coupling to phonons allows for  solitary charge-density waves propagating through the system \cite{cdw}, which might be probed by strong-field ionization or harmonic generation. In 2D and 3D, topological edge states exist in topological insulators or at interfaces in sandwiched materials \cite{topinsRevModPhys.82.3045,topins,topinsshortcourse}. Many interesting questions can be addressed in this context: How does the HHG efficiency in such materials depend on the laser polarization? As we have shown, the HHG yield depends not only on the band structure but also critically on the populations, which might open ways to control the HHG process electronically. On the other hand, one might control, e.g.,  the  spin currents along the edges of a topological insulator, with potential applications in laser-driven electronics. Another interesting question is whether one needs to shine with the laser on the edge of a sample to see the effect of edge states \cite{surfacebulkcorresp}%
. On the theoretical side, establishing a direct link between topological invariants and fingerprints of them in typical strong-field observables like HHG spectra would be desirable. That would, in principle, allow to probe the topology of matter in a single-shot strong-field experiment by all-optical means.

\bigskip

{\em Acknowledgments.} --- K.K.H. acknowledges support from the Villum-Kann Rasmussen (VKR) center of excellence QUSCOPE - Quantum Scale Optical Processes.

\bibliography{bibliography}

\end{document}